\documentclass[fleqn,12pt,twoside]{article}
\usepackage{espcrc1}

\newcommand{\sst}{\scriptscriptstyle}

\def \im {{\rm i}}

\title{An EFT for the weak $\Lambda N$ interaction
\thanks{Supported by 
EURIDICE HPRN--CT--2002--00311 (EU) and
DGICYT BFM2002--01868 (MCyT)}
}
\author{
\underline{A. Parre\~no}\thanks{Email address: assum@ecm.ub.es}
\address{Dep. ECM, Facultat de F\'{\i}sica,
U Barcelona, E-08028, Barcelona, Spain},
C. Bennhold\address{Center of Nuclear Studies, GWU,
Washington DC, 20052, USA},
B.R. Holstein\address{Dpt. of Physics-LGRT, University of Massachusetts,
Amherst, MA 01003, USA}
}

\date{today}
\begin{document}
\maketitle

\begin{abstract}

The nonleptonic weak $|\Delta S|=1$ $\Lambda N$ interaction, responsible
for the dominant, nonmesonic decay of all but the lightest hypernuclei, is
studied in the framework of an effective field theory. The long-range
physics is described through tree-level exchange of the SU(3) Goldstone
bosons ($\pi$ and $K$), while the short-range potential is parametrized
in terms of lowest-order contact terms
obtained from the most general non-derivative local four-fermion interaction.
Fitting to available weak hypernuclear decay
rates for $^5_\Lambda {\rm He}$, $^{11}_\Lambda {\rm B}$ and
$^{12}_\Lambda {\rm C}$ yields reasonable values for the low-energy constants.

\end{abstract}


\vspace*{0.5cm}
$\Lambda$-hypernuclei are bound systems composed of nucleons and one or 
more $\Lambda$ hyperons. 
For the past 50 years, these bound systems have been used to extend our
knowledge of both the strong and the weak baryon-baryon
interaction from the $NN$ case into the SU(3) sector.
Up to date there are no stable hyperon beams, which makes 
hypernuclear weak decay the only source of information about
the $|\Delta S|=1$ four-fermion (4P) interaction. Since the decay of medium to
heavy hypernuclei proceeds mainly via the $\Lambda N \to NN$ reaction, 
which does not conserve either parity, isospin or strangeness,
its study complements the weak $\Delta S=0$ $NN$ case which allows the 
study of only the parity-violating (PV) amplitudes, because the parity-conserving
(PC) signal is masked by the orders-of-magnitude stronger strong interaction
background.

Since the earliest hypernuclear studies it is well known that the 
one-pion exchange mechanism (OPE), which naturally explains the 
long-range part of the interaction, is able to fairly reproduce the 
total NonMesonic Decay (NMD) rate of hypernuclei, 
but not the partial rates, produced by the proton-induced 
mechanism ($\Gamma_p : \Lambda p \to np$) and by the  
neutron-induced mechanism ($\Gamma_n : \Lambda n \to nn$). 
The $\Lambda N$ mass difference, on the other hand, produces nucleons
with momenta around $\approx$ 420 MeV, suggesting that the short-range
part of the interaction cannot be neglected. 
These contributions have been described by:
1) the exchange of heavier mesons\cite{PRB97,DFHT96}
whose production thresholds are too high for the free $\Lambda$ decay;
2) an effective quark hamiltonian\cite{ISO00}; 
3) correlated 2$\pi$-exchanges in the form of $\sigma$ and $\rho$ 
mesons\cite{itonaga}; 4) $K$-exchange plus 
correlated and uncorrelated 2$\pi$-exchanges\cite{JOP01}.

The remarkable success of effective
field-theory techniques based on chiral expansions in
the SU(2) sector\cite{BBHPS,UvK99,ZPHR01,Ulf}
suggests extending this approach to the SU(3) realm,
even though stability of the chiral
expansion is less clear here due to the
significant degree of SU(3) symmetry breaking.  A well-known
example of the problems facing SU(3) chiral perturbation theory has 
been the prediction\cite{jenkins92,springer99,bh,borasoy03}
of the four PC p-wave amplitudes in the
weak nonleptonic decays of octet baryons, $Y \to N \pi$, with
$Y=\Lambda, \Sigma$ or $\Xi$.

The aim of the present contribution is to build a less model
dependent theory for the underlying $|\Delta S|=1$ $\Lambda N$ 
interaction governing hypernuclear decay, and see if 
a low order Effective Field Theory (EFT) can describe
the present available hypernuclear decay data.
Studies in this direction have already started~\cite{Jun,PBH04}.
In Ref.~\cite{Jun} a Fermi (V-A) interaction was added to the OPE mechanism to describe
the weak $\Lambda N \to NN$ transition. Ref.~\cite{PBH04} is a former
version of the present manuscript, the latter containing a few more results. 

In order to build such an EFT we shall allow for all possible contact terms 
in the four-baryon interaction Lagrangian, and fit the Low Energy Coefficients
(LEC), which size those contributions, to 
the available data in the appropriate energy range.
In contrast to the $NN$ case, however, the $\Lambda N \to NN$ transition
corresponds to an approximate energy release of $177$ MeV 
at threshold.
It is therefore not at all clear if low-energy expansions
can be carried out with any validity. In light of this threshold momentum 
value, \mbox{$|{\vec p}\ | \approx 417$ MeV}, 
it is thus reasonable to include the pion
\mbox{($m_\pi \approx 138$ MeV)} and the kaon \mbox{($m_K \approx 494$ MeV)} as
dynamical fields. Working within SU(3) also supports treating
pion and kaon on equal footing.
The last member of the SU(3) Goldstone-boson octet, the $\eta$,
is usually not included, since
the strong $\eta NN$ coupling is an order of magnitude
smaller than the strong $\pi NN$ and $K \Lambda N$ ones~\cite{etacoup}.
Therefore, in the present study we will describe the long-range
part of the weak $\Lambda N \to NN$ transition through pion and kaon exchanges,
while the short-range interaction will be parametrized through leading-order 
contact terms. Not considered here is the intermediate-range 2$\pi$-exchange.
Such a piece is two orders higher in the chiral expansion than the
corresponding single pion-exchange piece.  
The pion exchange is given by the following strong and weak Lagrangians:
\vspace*{-0.1cm}
\begin{eqnarray}
{\cal L}^{\rm S }_{\rm {\sst NN} \pi} &=& -\im \, g_{\rm{\sst NN} \pi} \,
\overline{\psi}_{\rm N}
\gamma_5 \gamma^\mu \, {\vec \tau} \cdot \partial_\mu \,
{\vec \phi}^{ \, \pi} \psi_{\rm N} \nonumber \\ 
{\cal L}^{\rm W }_{\rm {\sst \Lambda N} \pi} &=& - \im G_F m_\pi^2
\overline{\psi}_{\rm N}
(A_\pi + B_\pi \gamma_5 \gamma^\mu \,)
{\vec \tau} \,\cdot \partial_\mu \, {\vec \phi}^{\, \pi}
\psi_\Lambda \, \left( ^0_1 \right)  {\rm .}
\label{eq:pionlag}
\end{eqnarray}

\noindent Here, $G_F m_\pi^2= 2.21\times 10^{-7}$
is the weak coupling constant, $g_{\rm{\sst NN} \pi}=13.16$,
and the empirical constants
$A_\pi=1.05$ and $B_\pi=-7.15$, adjusted to the observables of the
free $\Lambda$ decay, determine
the strength of the parity-violating and parity-conserving
amplitudes, respectively.
The iso-spurion $\left( ^0_1 \right)$ is included
to enforce the empirical $\Delta I=1/2$ rule observed in the decay of a free $\Lambda$.
Performing a nonrelativistic reduction of the
resulting Feynman amplitude one obtains the OPE 
transition potential in momentum space:
\begin{eqnarray}
V_{\rm OPE} ({\vec q}\,) &=& - G_F m_\pi^2
\frac{g_{\rm{\sst NN} \pi}}{2 M_N} \left(
A_\pi + \frac{B_\pi}{2 \overline{M}}
{\vec \sigma}_1 \cdot {\vec q} \,\right) 
\, \frac{{\vec \sigma}_2 \cdot {\vec q}\,} {{\vec q}^{\; 2}+{m_\pi}^{\, 2}} \,
{\vec \tau}_1 \cdot {\vec \tau}_2 {\rm ,}
\label{eq:pion}
\end{eqnarray}
where ${\vec q}$ represents the momentum transfer directed towards the
strong vertex~\cite{PR01}, $M_N$ is the nucleon mass and
$\overline{M} = (M_N+M_\Lambda)/2$ is the average of the $N$ and $\Lambda$ 
masses.

The corresponding Lagrangians for the one-kaon-exchange (OKE) mechanism read:
\begin{eqnarray}
{\cal L}^{\rm S }_{\rm \sst{\Lambda N K}}&=&
- \im \, g_{\rm {\sst \Lambda N K}} \, \overline{\psi}_{\rm N}
\gamma_5 \, \gamma^\mu \, \partial_\mu \, \,
\phi^{\rm K} \psi_\Lambda \, \nonumber \\ 
{\cal L}^{\rm W }_{\rm \sst{NNK}}&=&
- \im \, G_F m_\pi^2 \, \left[ \overline{\psi}_{\rm N} \left( ^0_1 \right)
\,\,( C_{\rm \sst{K}}^{\rm \sst{PV}} +
C_{\rm  \sst{K}}^{\rm \sst{PC}}
\gamma_5 \, \gamma^\mu \, \partial_\mu \, \, )
(\phi^{\rm K})^\dagger
\psi_{\rm N} \right.
\nonumber \\
& & \left. + \, \overline{\psi}_{\rm N} \psi_{\rm N}
\,\,( D_{\rm \sst{K}}^{\rm\sst{PV}}
+ D_{\rm\sst{K}}^{\rm\sst{PC}}
\gamma_5 \, \gamma^\mu \, \partial_\mu \, \, )
(\phi^{\rm K})^\dagger \,\,
\left( ^0_1 \right) \right] {\rm .}
\label{eq:kaon}
\end{eqnarray}

\noindent The weak $C_K$ and $D_K$ coefficients are related
to the coupling constants at the ${\rm p}\overline{{\rm n}}{\rm
K}^+$ and ${\rm p}\overline{{\rm p}}{\rm K}^0$ vertices,
respectively. Like the $NN\pi$ coupling constant, the $\Lambda N
K$ and $\Sigma N K$ ones are a fundamental input
into our calculation. Unlike the $NN\pi$ coupling, however, their
values are less well known, with $g_{\rm{\sst \Lambda N K}} = 13 -
17 $ and $g_{\rm{\sst \Sigma N K}} = 3-6 $.  Here, we choose the
values given by the Nijmegen Soft-Core model f, NSC97f~\cite{nij99},
$g_{\rm{\sst \Lambda N K}} = -17.66$ and $g_{\rm{\sst \Sigma N K}} = -5.38$.
The corresponding nonrelativistic OKE potential is analogous to
Eq. (\ref{eq:pion}), but with the replacements~\cite{PRB97}:
\begin{eqnarray}
g_{\rm{\sst NN \pi}} &\to& g_{\rm{\sst \Lambda N K}} {\rm ,} \, \,
\,\,
m_\pi \to m_{\rm {\sst K}} \nonumber \\
A_\pi \; {\vec \tau}_1 \cdot{\vec \tau}_2 &\to& \left( \frac{
C^{\rm\sst{PV}}_{\rm\sst{K}}}{2} +
D^{\rm\sst{PV}}_{\rm\sst{K}} + \frac{
C^{\rm\sst{PV}}_{\rm\sst{K}}}{2}
{\vec \tau}_1 \cdot {\vec \tau}_2 \,\right) \frac{{\overline M}}{M_N} 
\nonumber \\
B_\pi \; {\vec \tau}_1 \cdot{\vec \tau}_2 &\to& \left( \frac{
C^{\rm\sst{PC}}_{\rm\sst{K}}}{2} +
D^{\rm\sst{PC}}_{\rm\sst{K}} + \frac{
C^{\rm\sst{PC}}_{\rm\sst{K}}}{2}
{\vec \tau}_1 \cdot {\vec \tau}_2 \right) \, {\rm .}
\end{eqnarray}

\noindent Of course, the weak NNK
couplings are not accessible
experimentally, so we obtain numerical values by making
use of SU(3) and chiral algebra considerations~\cite{PRB97,DFHT96}.
In addition, our OPE and OKE potentials will be regularized by using monopole
form factors at each vertex~\cite{PR01}. We note that all the strong model 
dependent ingredients used in the present calculation (as cut-off parameters
or strong coupling constants)
have been taken from the NSC97f interaction model~\cite{nij99}.

\vspace*{-0.3cm}
\begin{table}[hbt]
\caption{$\Lambda N \to NN$ partial waves.}
\begin{center}
\begin{tabular}{lccc}
\hline
{\rm partial \,\, wave} & {\rm operator} & {\rm order} & I \\
\hline
$a: ^1S_0 \to ^1S_0$ & ${\hat 1}, \, {\vec \sigma_1} {\vec \sigma_2}$ 
& $1$ & $1$ \\
$b: ^1S_0 \to ^3P_0$ & $({\vec \sigma_1} - {\vec \sigma_2}) {\vec q} {\rm ,} 
\,\, ({\vec \sigma_1} \times {\vec \sigma_2}) {\vec q}$ & $q/M_N$ & $1$ \\
$c: ^3S_1 \to ^3S_1$ & ${\hat 1}, {\vec \sigma_1} {\vec \sigma_2}$ & $1$ & 0 \\
$d: ^3S_1 \to ^1P_1$ & $({\vec \sigma_1} - {\vec \sigma_2}) {\vec q} {\rm ,} 
\,\, ({\vec \sigma_1} \times {\vec \sigma_2}) {\vec q} $ & $q/M_N$ & $0$ \\
$e: ^3S_1 \to ^3P_1$ & $({\vec \sigma_1} + {\vec \sigma_2}) {\vec q}$ 
& $q/M_N$ & $1$ \\
$f: ^3S_1 \to ^3D_1$ & $({\vec \sigma_1} \times {\vec q}) ({\vec \sigma_2} 
\times {\vec q})$ & $q^2/{M_N}^2$ & $0$ \\
\hline
\end{tabular}
\end{center}
\end{table}

\vspace*{-0.8cm}
If no model is assumed, the low energy $\Lambda N \to N_1 N_2$ process can be
parametrized through the 6 partial waves listed in Table \setcounter{footnote}{0}I~\footnote{
This illustrative argument is only valid for the $L=0$ part of the $\Lambda N$
wave function.}.
The PC $a$ and $c$ transitions can only be produced
by the ${\hat 1} \, \cdot \, \delta^3({\vec r} \ )$ and
${\vec{\sigma_1} \cdot {\vec \sigma_2}} \, \cdot \, \delta^3({\vec r} \ )$ operators,
where $\delta^3({\vec r} \ )$ represents the contact interaction.
The PV $b$ and $d$ transitions proceed through
the combination of the spin-nonconserving operators:
$({\vec \sigma_1} - {\vec \sigma_2})$ and 
$\im ({\vec \sigma_1} \times {\vec \sigma_2})$ with the following operators:
$\{ \vec{p}_1 - \vec{p}_2 \, , \, \delta^3({\vec r} \ ) \, \}$
and $[ \, \vec{p}_1 - \vec{p}_2\, , \, \delta^3({\vec r} \ )\,  ]$,
where $\vec{p}_i$ is the derivative operator acting on the "i{\it th}" particle
\footnote{Note that we are assuming
that ${\vec p}_1 - {\vec p}_2$ is small enough to disregard higher powers of
the derivative operators $ \vec{p}_1 - \vec{p}_2 $.}.
The $e$ transition is allowed by the combination of the spin-conserving operator
$({\vec \sigma_1} + {\vec \sigma_2})$ with the previous commutator and 
anti-commutator,
while only two-derivative operators can produce the last
(tensor) transition.
In order to write the general four-fermion interaction we will 
retain those terms depending on the momentum transfer 
${\vec q} = {\vec p}_\Lambda - {\vec p}_{N_1} = {\vec p}_{N_2} - {\vec p}_N
= {\vec k}_i - {\vec k}_f$, with ${\vec k}_i$ (${\vec k}_f$) being the initial 
(final) relative momentum in the c.m. frame~\footnote{This is a reasonable 
assumption for the 
weak $\Lambda N \to N_1 N_2$ process, where the
two particles in the initial state (bound in a hypernucleus) have very low
momentum, specially when compared to the momentum of the two outgoing nucleons. 
That means that we can neglect terms containing ${\vec k}_i$ 
in the expansion, and at the same time approximate 
${\vec k}_f$ by $- {\vec q}$.}. The 4P potential reads
(in units of $G_F = 1.166 \times 10^{-11}$ MeV$^{-2}$):
\begin{eqnarray}
V_{4P} ({\vec q}) &=&
C_0^0 + C_0^1 \; {\vec \sigma}_1 \cdot{\vec \sigma}_2 \,\,
+ C_1^0 \; \displaystyle\frac{{\vec \sigma}_1 \cdot{\vec q}}
{2 {\overline M}}
+ \,C_1^1 \; \displaystyle\frac{{\vec \sigma}_2\cdot {\vec q}}{2 M}
+ {\rm i} \, C_1^2 \; \displaystyle\frac{({\vec \sigma}_1 \times {\vec \sigma}_2)
\; \cdot {\vec q}}{2 \tilde{M}}
\\
&+& C_2^0 \; \displaystyle\frac{{\vec \sigma}_1 \cdot{\vec q} \;
{\vec \sigma}_2 \cdot{\vec q}}{4 M {\overline M}} +
C_2^1 \; \displaystyle\frac{{\vec \sigma}_1\cdot {\vec \sigma}_2 \; {\vec q}^{\; 2}}
{4 M {\overline M}} +
C_2^2 \; \displaystyle\frac{{\vec q}^{\,2}}{4 M \tilde{M}}
\nonumber
\end{eqnarray}
where $\tilde{M}= \displaystyle\frac{3 M+M_\Lambda}{4}$ is a weighted average
of $N,\Lambda$ masses and $C_i^j$
is the j$th$ LEC at i$th$ order. While the form of the
contact terms is model independent, the size of the coefficients 
depends on how the theory is formulated,
and they are expected to be of the order of the other couplings in the problem.
These couplings provide a very simple representation of the short
distance contributions to the process at hand.  In a complete model,
they would be represented by specific dynamical contributions, such as
$\rho,\omega, etc.$-exchange.  However, we eschew the temptation to be
more specific---in fact this generality is one of the strengths of our
approach.  We evaluate the coefficients purely phenomenologically and
leave the theoretical interpretation of the pieces to future
investigations.  Of course, the specific size of such coefficients depends upon
the chiral order to which we are working.  However, if the expansion
is convergent, then the values of these effective couplings should be
relatively stable as NLO or higher effects are included.

To reduce the number of free parameters we use power
counting, discarding operators of order $q^2/{M_N}^2$.
To obtain the 4P potential in configuration space we must Fourier transform
$V_{\rm 4P} (\,{\vec q\,})$, smearing the resulting delta functions
by using
a normalized Gaussian form for the 4-fermion contact
potential,
$f_{ct}(r)= e^{- \frac{r^2}{\delta^2}}/( \delta^3 \pi^{3/2})$,
where $\delta$ is taken to be of the order of a typical vector-meson range,
$\delta \sim \sqrt{2} m_\rho^{-1} \approx 0.36$ fm.
The leading order $V_{\rm 4P} (\,{\vec r}\,)$ potential for both PV and PC
terms can then be written as:
\begin{eqnarray}
V_{4P} ({\vec r}) &=&
\left\{C_0^0 + C_0^1 \; {\vec \sigma}_1 \cdot{\vec \sigma}_2
+ \displaystyle\frac{2 r}{\delta^2}
\left[ C_1^0 \; \displaystyle\frac{{\vec \sigma}_1 \cdot{\hat r}}{2 {\overline M}}
+ C_1^1 \; \displaystyle\frac{{\vec \sigma}_2 \cdot{\hat r}}{2 M}
+ C_1^2 \; \displaystyle\frac{({\vec \sigma}_1 \times {\vec \sigma}_2)
\cdot {\hat r}}{2 {\tilde{M}}} \right]\right\} \nonumber \\
&\times& f_{ct} (r)
\; \times \left[
C_{IS} \, {\hat 1} + C_{IV} \,
{\vec \tau_1}\cdot {\vec \tau}_2
\right] \, {\rm ,}
\label{eq:4ppotr}
\end{eqnarray}
where the last factor represents the
isospin part of the 4-fermion interaction. Note that
we only allow for $\Delta I=1/2$ transitions.


It is well known that the high momentum transferred in the $\Lambda N \to NN$
reaction makes this process sensitive to the short range
physics which is characterized by our contact coefficients.
Moreover, since the $|\Delta S|=1$ reaction takes place in a finite nucleus,
extracting information of the elementary weak two-body
interaction requires a careful investigation of the many-body nuclear
effects present in the hypernucleus.
The nonmesonic decay rate is written as:
$$\Gamma_{\rm nm} = \int \frac{d^3 k_1}{(2\pi)^3}
\int \frac{d^3 k_2}{(2\pi)^3}
\sum\limits_{\stackrel{M_J \{R\}}{\{1\} \{2\}}}
(2\pi) \,
\delta(M_H-E_R-E_1-E_2)
\frac{1}{(2J+1)}
\mid {{\cal M}_{fi} }\mid^2 {\rm ,}
$$
where $k_1$ and $k_2$ represent the momenta of the two outgoing nucleons,
$J$ and $M_J$ the total spin and spin projection of the initial hypernucleus,
$R$ the quantum numbers of the residual nucleus, $M_H$ the mass of the 
hypernucleus, and ${{\cal M}_{fi} }$ the hypernuclear transition amplitude,
given by $ {\cal M}_{fi}
\sim \langle {\vec k}_1 m_1 {\vec k}_2 m_2; \Psi_{\rm R}^{A-2}
\mid {\hat O}_{\Lambda {\rm N} \to {\rm NN}}
\mid ^A_{\Lambda} Z \rangle {\rm .} $
\vspace*{-0.4cm}
\begin{table}[hbt]
\caption{Experimental data for the decay of $^5_\Lambda$He, $^{11}_\Lambda$B
and $^{12}_\Lambda$C. The values underlined are the ones used in our fits.}
\begin{center}
\begin{tabular}{l|c|c|c|c}
\hline
 & $\Gamma$ & $\Gamma_{\rm n}/\Gamma_{\rm p}$ & $\Gamma_{\rm p}$ & ${\cal A}$ \\\hline
$^5_\Lambda$He &
\underline{$0.41 \pm 0.14$}~\cite{Szy91} & $0.93\pm 0.55$~\cite{Szy91} &
\underline{$0.21 \pm 0.07$}~\cite{Szy91} & \underline{$0.24 \pm 0.22$}~\cite{Ajim00}  \\
 & \underline{$0.50 \pm 0.07$}~\cite{No95}  & $1.97 \pm 0.67$~\cite{No95} & & \\
 & & \underline{$0.50 \pm 0.10$}~\cite{kek02} & & \\
\hline
$^{11}_\Lambda$B &
\underline{$0.95 \pm 0.14$}~\cite{No95} &
\underline{$1.04^{+0.59}_{-0.48}$}~\cite{Szy91}&
$0.30^{+0.15}_{-0.11}$~\cite{No95} & $-0.20\pm 0.10$~\cite{Aj92} \\
 & & $2.16 \pm 0.58^{+0.45}_{-0.95}$~\cite{No95} & &  \\
 & & $0.59^{+0.17}_{-0.14}$~\cite{Mo74}  & &  \\
\hline
$^{12}_\Lambda$C &
  \underline{$0.83\pm 0.11$}~\cite{Bhang} & $1.33^{+1.12}_{-0.81}$~\cite{Szy91} &
$0.31^{+0.18}_{-0.11}$~\cite{No95} & $-0.01\pm 0.10$~\cite{Aj92} \\
 & \underline{$0.89 \pm 0.15$}~\cite{No95} &
$1.87 \pm 0.59^{+0.32}_{-1.00}$~\cite{No95} & & \\
 & \underline{$1.14\pm 0.2$}~\cite{Szy91}&
$0.59^{+0.17}_{-0.14}$~\cite{Mo74}  & &  \\
 & & \underline{$0.87 \pm 0.23$}~\cite{Ha02b} & & \\
\hline
\end{tabular}
\end{center}
\end{table}

\vspace*{-0.8cm}
In the present calculation, we use a shell-model for the initial hypernucleus
and assume a weak coupling scheme for the hyperon (the $\Lambda$ particle 
will only couple the ground state core wave function).
Moreover, using coefficients of fractional 
parentage we can decouple the weakly interacting nucleon from the
antisymetrized core wave function, leaving also an antisymmetric wave function
for the residual system. 
The single-particle $\Lambda$ and N orbits
are taken to be solutions of harmonic oscillator mean field potentials
with parameters adjusted to experimental separation energies
and charge form factor of the hypernucleus under study.
The strong hyperon-nucleon interaction at short distances,
absent in mean-field models, is accounted for by replacing the
mean-field two-particle $\Lambda N$ wave function by
a correlated~\cite{sitges} one inspired in 
a microscopic finite-nucleus $G$-matrix
calculation ~\cite{Ha93} which uses the soft-core and hard-core
Nijmegen models ~\cite{NR77}.
The $NN$ wave function is obtained by solving the
Lippmann-Schwinger ($T$-matrix) equation with the input of the
Nijmegen Soft Core NSC97f potential model (details of the calculation
can be found in the Appendix of Ref. ~\cite{PR01}). 

\begin{table}[hbt]
\caption{Results for the weak decay observables,
when a fit to the $\Gamma$ and $n/p$ for
$^5_\Lambda {\rm He}$, $^{11}_\Lambda {\rm B}$ and
$^{12}_\Lambda {\rm C}$ is performed.
The values in parentheses have been obtained including
$\alpha_\Lambda (^5_\Lambda {\rm He})$ in the fit.
        }
\begin{center}
\begin{tabular}{|l|c|c|c|c|c|}
\hline
 & $\pi$ & $+ K$ & $+$ LO PC & $+$ LO PV & EXP: \\
\hline
$\Gamma (^5_\Lambda {\rm He})$ & $0.46$ & $0.29$ & $0.44$ & $0.44$ ($0.44$)
 & $0.41 \pm 0.14$~\cite{Szy91}, ~$0.50 \pm 0.07$~\cite{No95}  \\
$n/p (^5_\Lambda {\rm He})$ & $0.09$ & $0.51$ & $0.56$ & $0.54$ ($0.54$)
& $0.93\pm 0.55$~\cite{Szy91}, ~$0.50 \pm 0.10$~\cite{kek02}  \\
$\alpha_\Lambda (^5_\Lambda {\rm He})$ & $-0.23$ & $-0.54$
& $-0.72$ & $-0.33$ ($0.24$) & $0.24 \pm 0.22$~\cite{Ajim00} \\
\hline
$\Gamma (^{11}_\Lambda {\rm B})$ & $0.67$ & $0.43$
& $0.87$ & $0.87$ ($0.87$)
 &  $0.95  \pm 0.14$~\cite{No95}  \\
$n/p (^{11}_\Lambda {\rm B})$ & $0.11$ & $0.45$
 & $0.82$ & $0.99$ ($0.99$)
 & $1.04^{+0.59}_{-0.48}$~\cite{Szy91} \\
${\cal A} (^{11}_\Lambda {\rm B})$ & $-0.11$ & $-0.24$
& $-0.20$ & $-0.02$ ($0.12$)
 & $-0.20 \pm 0.10$~\cite{Aj92}  \\
\hline
$\Gamma (^{12}_\Lambda {\rm C})$ & $0.80$ & $0.49$
& $0.95$ & $0.93$ ($0.93$)
 & $1.14\pm 0.2$~\cite{Szy91}, $0.89 \pm 0.15$~\cite{No95} \\
 & & & & & $0.83\pm 0.11$~\cite{Bhang} \\
$n/p (^{12}_\Lambda {\rm C})$ & $0.09$ & $0.36$
& $0.64$ & $0.82$ ($0.81$)
 & $0.87 \pm 0.23$~\cite{Ha02b} \\
${\cal A} (^{12}_\Lambda {\rm C})$ & $-0.03$ & $-0.06$
& $-0.05$ & $-0.006$ ($0.03$)
 & $-0.01 \pm 0.10$~\cite{Aj92}  \\
\hline
${\hat \chi^2}$ & & & $0.98$ & $1.49$ ($1.12$) & \\
\hline
\end{tabular}
\end{center}
\end{table}

\vspace*{-0.8cm}
Before we start with the discussion of the results, we should make a remark on the data.
One might wonder if there can be only three independent data points in the 
nonmesonic decay: the proton-induced and neutron-induced rates $\Gamma_p$ and 
$\Gamma_n$, and the asymmetry $\cal {A}$ (associated with the proton-induced 
decay), relating observables from one hypernucleus to another through 
hypernuclear structure coefficients.  While one may
indeed expect measurements from different p-shell hypernuclei, say, A=12 and 16,
to provide the same constraint, the situation is different when including data
from s-shell hypernuclei like A=5. For the latter, the initial $\Lambda N$ pair
 can only be in a relative s-state, while for the former, relative p-states are
 allowed as well.  We therefore include in our fits the data shown in Table 2 
from the A=5,11 and 12 hypernuclei. 
Only more recent measurements from the
last 12 years were used, however, we excluded those recent data of the ratio
$\Gamma_n$/$\Gamma_p = n/p$ whose
error bars were larger than 100$\%$.  We also have taken the data at
face value and have
not applied any corrections due to, e.g., the two-nucleon induced
mechanism~\cite{2N}. Given the sizable error bars
of the data, this omission does not change any of our conclusions.
New exclusive measurements will hopefully
be able to verify the magnitude of such contributions.

No parameters were fitted for the results with only
$\pi$ and $K$ exchange, shown in \mbox{Table 3.}  
As has been known for a long time, $\pi$ exchange
alone describes reasonably well the observed total rates, while dramatically
underestimating the $n/p$ ratio.
Incorporation of kaon exchange gives a destructive interference between both
mechanisms (OPE and OKE) in the PC amplitudes,
while the interference is constructive in the PV ones.
The tensor PC channel dominates the proton-induced rate while it is absent
in the $L=0$ neutron-induced one. As a consequence, 
~$n/p$ is enhanced by about a factor of five, within reach of the lower
bounds of the experimental
 measurements, and the total rate underpredicts the observed value
by about a factor of two. It also leads to
values for the asymmetry that are close to experiment for the p-shell
hypernuclei, but far off for A=5~\footnote{Note that 
$\alpha_\Lambda$ stands for the intrinsic $\Lambda$ asymmetry parameter, characteristic
of the elementary \mbox{$\Lambda N \to NN$} reaction. 
For helium, this quantity can be extracted directly from
experiment without the model dependent input of the $\Lambda$ (or hypernuclear) 
polarization~\protect\cite{Ajim00}.}. Since the contributions of both
$\eta$-exchange and two-pion exchange are negligible, these discrepancies
illustrate the need for short-range physics.
\vspace*{-0.4cm}
\begin{table}[hbt]
\caption{LEC coefficients corresponding to the LO
calculation.
The values in parentheses have been obtained including
$\alpha_\Lambda (^5_\Lambda {\rm He})$ in the fit.}
\begin{center}
\begin{tabular}{|l|c|c|}
\hline
 & $+$ LO PC & $+$LO PV  \\
\hline
$C_0^0$ & $-1.46 \pm 0.44$ & $-1.12 \pm 0.52$
($-0.95 \pm 0.31$)\\
$C_0^1$ & $-0.85 \pm 0.27$ & $-0.99 \pm 0.74$
($-0.53 \pm 0.24$)\\
$C_1^0$ & $---$ & $-5.71 \pm 4.16$
($-4.99 \pm 1.62$)\\
$C_1^1$ & $---$  & $2.95 \pm 2.90$
($2.71 \pm 1.91$)\\
$C_1^2$ & $---$ & $-6.56 \pm 1.78$
($-5.24 \pm 1.93$)\\
$C_{IS}$ & $4.85 \pm 1.45$ & $4.69 \pm 0.11$
($5.97 \pm 0.71$)\\
$C_{IV}$ & $1.43 \pm 0.45$ & $1.25 \pm 0.11$
($1.59 \pm 0.21$)\\
\hline
\end{tabular}
\end{center}
\end{table}

\vspace*{-0.8cm}
Allowing contact terms of order unity (leading-order PC operators)
to contribute leads to four free parameters,
$C_0^0$, $C_0^1$, $C_{IS}$ and $C_{IV}$.
Data on the total and partial decay rates for all three hypernuclei are
included in the fit, but no asymmetry measurements.
The inclusion of the contact terms roughly doubles the values for the total
decay rates, thus restoring agreement with experiment.
The impact on the $n/p$ ratio is noteworthy: the value for
$^5_\Lambda {\rm He}$ increases by 10$\%$ while the
$n/p$ ratios for $^{11}_\Lambda {\rm B}$ and $^{12}_\Lambda {\rm C}$ almost
double. This is an example of the differing impact certain operators can
have for s- and p-shell hypernuclei. The effect on the asymmetry is opposite,
almost no change for A=11 and 12, but a 30$\%$ change for A=5.
The magnitudes of the four parameters, $C_0^0$, $C_0^1$, $C_{IS}$
and $C_{IV}$, listed in Table~4,
are all around their natural size of unity, with the exception of
$C_{IS}$ which is a factor of five or so larger. Note the substantial
error bars on all the parameters, reflecting the uncertainties in the measurements.

Three new parameters are admitted when we allow the leading-order PV terms
(of order $q/M_N$) to
contribute with the coefficients $C_1^0$, $C_1^1$, and $C_1^2$.
As shown in Table~4, the parameters for the PV contact terms are
larger than the ones for the PC terms, and in fact, compatible
with zero. Including the three new parameters does not
substantially alter the previously fitted ones, thus supporting
the validity of our expansion. Regarding their impact on the
observables, the PV contact terms barely modify the total and
partial rates but significantly affect the asymmetry, as one would
expect for an observable defined by the interference between PV
and PC amplitudes. 
The calculated asymmetry considerably decreases in size for all three
hypernuclei, giving p-shell values close to zero, but still negative.
In order to further understand this
behavior, we have performed a number of fits including the
asymmetry data points of either $^5_\Lambda {\rm He}$ or
$^{11}_\Lambda {\rm B}$ or both. Tables~3 and 4 display the
result of one of those fits. Inclusion of the $^5_\Lambda {\rm
He}$ (intrinsic $\Lambda -$) asymmetry helps in constraining the values of two of the LO
PV parameters. We find that the two present experimental values
for A=5 and A=11 cannot be fitted simultaneously with this set of
contact terms. Future experiments will have to settle this issue.

\vspace*{-0.4cm}
\begin{table}[hbt]
\caption{Sensitivity of the calculation to the strong
interaction model. The numbers in the left (right) column
use final $NN$ wave functions obtained from the NSC97f (NSC97a)
interaction\protect\cite{nij99}. The 
$\alpha_\Lambda (^5_\Lambda {\rm He})$ has been included in the fit.}
\begin{center}
\begin{tabular}{|l|c|c|l|c|c|}
\hline
\multicolumn{3}{|c|}{$\pi + K + \rm{LO} \; \rm{PC} + \rm{LO} \; \rm{PV} $} &
\multicolumn{3}{|c|}{$\pi + K + \rm{LO} \; \rm{PC} + \rm{LO} \; \rm{PV} $} \\
\hline
 & NSC97a & NSC97f & & NSC97a & NSC97f \\
\hline
$\Gamma (^{5}_\Lambda {\rm He})$ & $0.44$ & $0.44$ 
& $C_0^0$ & $-0.67 \pm 0.42$ & $-0.95 \pm 0.31$ \\
$n/p (^{5}_\Lambda {\rm He})$ & $0.55$ & $0.54$ & 
$C_0^1$ & $-0.34 \pm 0.32$ & $-0.53 \pm 0.24$ \\
$\alpha_\Lambda (^{5}_\Lambda {\rm He})$ & $0.24$ & $0.24$ & 
& &  \\
\hline
$\Gamma (^{11}_\Lambda {\rm B})$ & $0.86$ & $0.87$ & $C_1^0$ &
$-5.85 \pm 1.40$ & $-4.99 \pm 1.62$ \\
$n/p (^{11}_\Lambda {\rm B})$ & $0.95$ & $0.99$ & $C_1^1$ & 
$3.65 \pm 1.66$ & $2.71 \pm 1.91$ \\
${\cal A} (^{11}_\Lambda {\rm B})$ & $0.08$ & $0.12$ & $C_1^2$ & 
$-6.47 \pm 1.64$ & $-5.24 \pm 1.93$ \\
\hline
$\Gamma (^{12}_\Lambda {\rm C})$ & $0.94$ & $0.93$ & $C_{IS}$ & 
$ 5.78 \pm 0.86$ & $5.97 \pm 0.71$ \\
$n/p (^{12}_\Lambda {\rm C})$ & $0.77$ & $0.81$ & $C_{IV}$ & 
$ 1.65 \pm 0.27$ & $1.59 \pm 0.21$ \\
${\cal A} (^{12}_\Lambda {\rm C})$ & $0.02$ & $0.03$ & & & \\
\hline
${\hat \chi^2}$ & $1.19$ & $1.12$ &  &  &  \\
\hline
\end{tabular}
\end{center}
\end{table}

\vspace*{-0.9cm}
We have also performed fits allowing a contribution from an isospin
$\Delta I = 3/2$ transition operator in Eq.~6. 
The resulting fit shifts strength from the
isoscalar contribution to the new $\Delta I = 3/2$ one, leaving the
other parameters unchanged.
In any case, as shown in Table~3,
we can clearly get an excellent fit to all observables without such
transitions while obtaining constants of reasonable size.
Table~5 demonstrates that our conclusions are basically independent of the
model we use for the strong force used for describing Final State Interactions 
in the transition. Employing $NN$ wave functions that are obtained with
either the NSC97f or the NSC97a model, one can perfectly fit the total and partial
rates, as well as the helium asymmetry, while the predicted $p-$shell asymmetries 
show a variation range of $50 \%$. The constants can easily absorb the
changes (the largest change affects the LO PC $C_0^1$, which becomes compatible 
with zero when the NSC97a model is used) and remain compatible within their error bars. 
Similarly, the results in Table~6 show the insensitivity of the predicted
observables to realistic values of the $\delta$ range used to smear the delta 
function in the contact potential. 

\vspace*{-0.4cm}
\begin{table}[hbt]
\caption{Sensitivity of the calculation
to the 4-fermion contact range $\delta$ in the smeared delta function. The 
$\alpha_\Lambda (^5_\Lambda {\rm He})$ is included in the fit.}
\begin{center}
\begin{tabular}{|l|c|c|c|}
\hline
 & $\delta \approx 0.3 \rm{fm}$ & $\delta \approx 0.36 \rm{fm}$ &
$\delta \approx 0.4 \rm{fm}$ \\
\hline
${\cal A} (^{11}_\Lambda {\rm B})$ & $0.12$ & $0.12$ & $0.12$ \\
\hline
${\cal A} (^{12}_\Lambda {\rm C})$ & $0.03$ & $0.03$ & $0.03$  \\
\hline
$C_0^0$ & $-1.81 \pm 0.43$ & $-0.95 \pm 0.31$ & $-0.69 \pm 0.27$ \\
$C_0^1$ & $-1.03 \pm 0.37$ & $-0.53 \pm 0.24$ & $-0.37 \pm 0.20$ \\
$C_1^0$ & $-7.10 \pm 2.38$ & $-4.99 \pm 1.62$ & $-4.13 \pm 1.35$ \\
$C_1^1$ & $4.00 \pm 2.93$ & $2.71 \pm 1.91$ & $2.15 \pm 1.56$ \\
$C_1^2$ & $-7.44 \pm 3.01$ & $-5.24 \pm 1.93$ & $-4.33 \pm 1.56$ \\
$C_{IS}$ & $6.43 \pm 0.60$ & $5.97 \pm 0.71$ & $5.69 \pm 0.79$ \\
$C_{IV}$ & $1.82 \pm 0.19$ & $1.59 \pm 0.21$ & $1.44 \pm 0.23$ \\
\hline
${\hat \chi^2}$ & $1.12$ & $1.12$ & $1.13$ \\
\hline
\end{tabular}
\end{center}
\end{table}
\vspace*{-0.4cm}


\vspace*{-0.7cm}
In conclusion, we have studied the nonmesonic weak decay using an
Effective Field Theory
framework for the weak interaction.  The long-range components were
described with pion and
kaon exchange, while the short-range part is parametrized in
leading-order PV and PC contact terms. 
We find coefficients of natural size with significant error bars, reflecting the level of
experimental uncertainty.  
We found a large contribution from an isoscalar, spin-independent
central operator. 
There is no indication of any contact terms violating the $\Delta I$ = 1/2 rule.
In this study we have not speculated on the dynamical origin of these contact 
contributions but our aim
was to ascertain their size and verify the validity of the EFT framework for the weak 
decay.
The next generation of data from recent high-precision weak decay experiments 
currently under analysis
holds the promise to provide much improved constraints for studies of this nature.

\end{document}